\documentclass[12pt]{article}
\usepackage{times}
\usepackage{geometry}
\usepackage[utf8]{inputenc}
\usepackage{enumitem,amssymb}
\usepackage{ragged2e}

\usepackage{graphicx}
\usepackage{url}
\usepackage{color}
\usepackage{natbib}
\usepackage{aas_macros}
\usepackage{titlesec}
\titleformat{\section}
  {\normalfont\fontsize{14}{0}\bfseries}{\thesection}{0.5em}{}
\def\arcsec{$^{\prime\prime}$}

\newlist{thematic}{itemize}{8}
\setlist[thematic]{label=$\square$}
\usepackage{pifont}

\begin{document}
\raggedright
\huge
Astro2020 Science White Paper %\linebreak

%%% MLG Quick Links
%%% Astro2020 Overview http://sites.nationalacademies.org/SSB/CurrentProjects/SSB_185159
%%% Astro2020 Formatting http://sites.nationalacademies.org/cs/groups/ssbsite/documents/webpage/ssb_190176.pdf

\bigskip
%%% Possible Titles
% Uncharted Frontiers of Transient Discovery and Follow-Up: An ELT \& LSST Perspective
% Uncharted Frontiers of Transient Discovery and Follow-Up: ELT \& LSST Synergy
Discovery Frontiers of Explosive Transients: An ELT \& LSST Perspective %\linebreak
\normalsize

\medskip
\noindent \textbf{Thematic Areas:} \hspace*{60pt} $\square$ Planetary Systems \hspace*{10pt} $\square$ Star and Planet Formation \hspace*{20pt}\linebreak
$\boxtimes$ Formation and Evolution of Compact Objects \hspace*{31pt} $\boxtimes$ Cosmology and Fundamental Physics \linebreak
  $\boxtimes$ Stars and Stellar Evolution \hspace*{1pt} $\square$ Resolved Stellar Populations and their Environments \hspace*{40pt} \linebreak
  $\square$    Galaxy Evolution   \hspace*{45pt} $\square$             Multi-Messenger Astronomy and Astrophysics \hspace*{65pt} \linebreak
  
\textbf{Principal Author:} 
Melissa L. Graham, University of Washington, mlg3k@uw.edu

%Name:
% \linebreak						
%Institution:
% \linebreak
%Email:
% \linebreak
%Phone:  
% \linebreak
 
\medskip
\textbf{Co-authors:} 
Danny Milisavljevic (Purdue University); % convener
Armin Rest (Space Telescope Science Institute); % echoes lead
J.~Craig Wheeler (University of Texas at Austin); % Spol lead
Ryan Chornock (Ohio University); % convener
Raffaella Margutti (Northwestern University); % convener
Jeonghee Rho (SETI Institute); %early KSP contributor
Chien-Hsiu Lee (National Optical Astronomy Observatory); % comments and suggestions
Sung-Chul Yoon (Seoul National University); % SPol text
% Maryam Modjaz (New York University); % SPol text
Charles~D. Kilpatrick (University of California Santa Cruz); % echoes text and data
Gautham Narayan (Space Telescope Science Institute); % echoes text and data
Nathan Smith, G. Grant Williams (University of Arizona); % Smith = echoes text and data; Williams = SPol contributions
Niharika Sravan (Purdue University); % text edits
Philip Cowperthwaite (Carnegie Observatories); % cosigner
Deanne Coppejans, Giacomo Terreran, Adriano Baldeschi (Northwestern University); % cosigners
V. Zach Golkhou (University of Washington); %cosigner
and Sumner Starrfield (Arizona State University). %cosigner

\medskip
\textbf{Abstract:} The Large Synoptic Survey Telescope (LSST) will open a discovery frontier for faint and fast transients with its ability to detect variable flux components down to $\sim$24.5 mag in a $\sim$30 second exposure. Spectroscopic follow-up of such phenomena --- which are necessary for understanding the {\em physics} of stellar explosions --- can require a rapid response and several hours with a 8-10m telescope, making it both expensive and difficult to acquire. The future Extremely Large Telescopes (ELTs) would be able to provide not only spectroscopy but capabilities such as spectropolarimetry and high-resolution diffraction-limited imaging that would contribute to future advances in our physical understanding of stellar explosions. {\bf In this white paper we focus on several specific scientific impacts in the field of explosive transient astrophysics that will be generated by the combination of LSST's discovery abilities and ELTs' follow-up capacities.} First, we map the uncharted frontier of discovery phase-space in terms of intrinsic luminosity and timescales for explosive transients, where we expect the unexpected (\S~\ref{sec:phase}). We then focus on six areas with open science questions for known transients: the progenitors of thermonuclear supernovae (SNe; \S~\ref{sec:snia}), mass loss prior to core collapse (\S~\ref{sec:cc}), asymmetry in stellar explosions (\S~\ref{sec:spol}), light echoes (\S~\ref{sec:echos}), high-$z$ transients (\S~\ref{sec:highz}), and strongly lensed SNe (\S~\ref{sec:lens}). We conclude with a brief discussion of the practical aspects of ELT \& LSST synergy (\S~\ref{sec:datainfra}). Five companion white papers discuss related transient astrophysics enabled by ELT: late-phase SN spectroscopy \citep{DannyWP}, multi-messenger astrophysics \citep{2019arXiv190304629C}, the first explosions \citep{2019arXiv190301569W}, tidal disruption events \citep{2019arXiv190301575W}, and SNe as dust producers \citep{RhoDust}.

\pagebreak
%%%%%%%%%%%%%%%%%%%%%%%%%%%%%%%%%%%%%%%%%%%%%%%%%%%
\section{Explosive Transient Phase-Space}\label{sec:phase}
\begin{figure}[ht!]
\centering
\includegraphics[width=15cm]{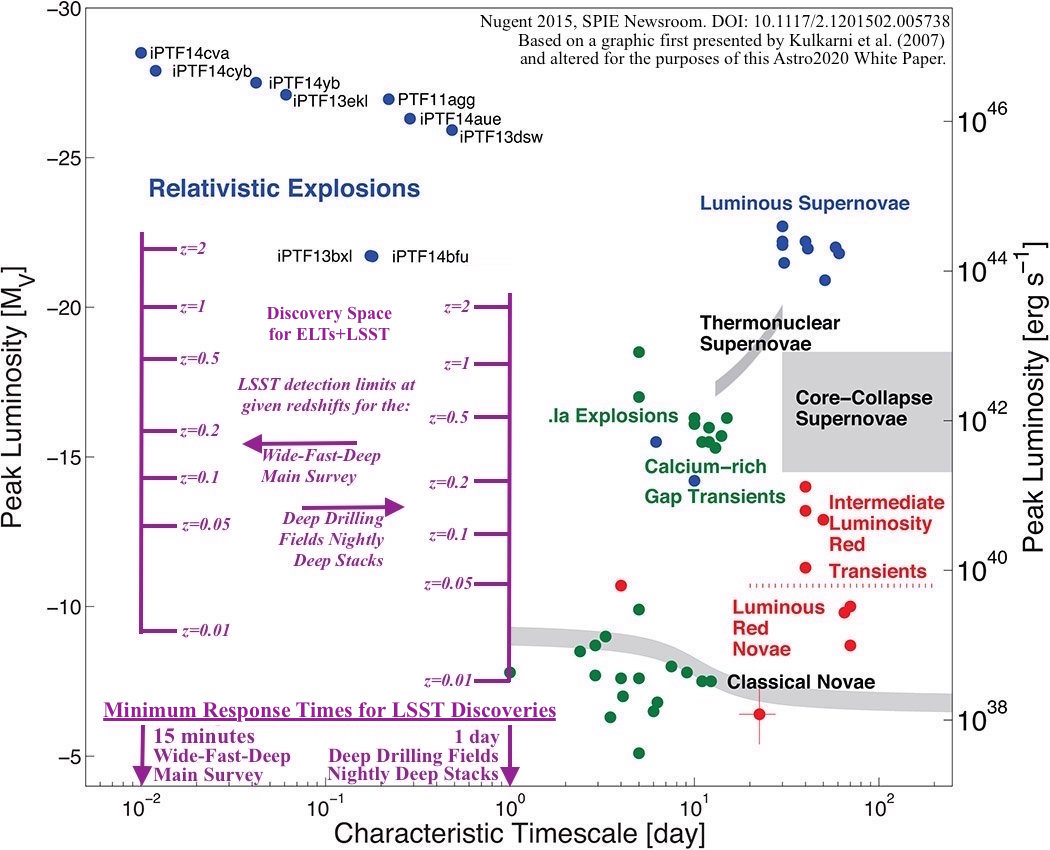}
\vspace{-5pt}
\caption{A phase-space diagram for explosive transients: peak intrinsic luminosity {\it vs.} characteristic timescale \citep{nugent_2015,2011PhDT........35K}.
The points and shaded regions represent transient types established by past surveys (e.g., the Palomar Transient Factory; PTF). In purple we sketch how ELT+LSST will fill the uncharted regions for timescales $<$1 day and luminosities ${\lesssim}-23$ mag. The LSST will release data on transients within 1 minute of readout and, e.g., the Thirty Meter Telescope (TMT) target-of-opportunity (ToO) acquisition time is $\sim$10 minutes \citep{TMTToO}: together they provide a minimum response time of $\sim$15 minutes. We use a limiting magnitude of $r{\sim}24$ mag for a $30$s LSST image and show the intrinsic magnitudes reached for a given redshift with the vertical scale at ${\sim}$15 min. The LSST will also perform deep drilling observations in small areas with a nightly stacked depth of $r{\sim}26$ mag, represented by the scale at $\sim$1 day. All LSST information is from \citet{2008arXiv0805.2366I}.}
\label{fig:1}
\end{figure}

%%% MLG quick links
%%% Ivezic+08 LSST: science drivers to design paper https://arxiv.org/pdf/0805.2366.pdf
\vspace{-5pt}
Fig. \ref{fig:1} shows a phase-space illustrating transient discoveries: the peak intrinsic luminosity {\it vs.} characteristic timescale. Some known transients are shown with points and shading, and the uncharted territories --- the blank space and beyond the borders --- are the discovery frontiers to be explored by LSST: intrinsically faint, short- or long-duration, and/or high-redshift objects. {\bf A physical understanding of transient phenomena at these discovery frontiers needs spectra, which will require the ELTs.} Not represented in Fig. \ref{fig:1} is the ELT-enabled discovery frontier provided by instrumental capabilities like high-resolution imaging and spectropolarimetry.
\vspace{-5pt}

\clearpage
%%%%%%%%%%%%%%%%%%%%%%%%%%%%%%%%%%%%%%%%%%%%%%%%%%%
\section{The Progenitor Scenario(s) of Thermonuclear Supernovae}\label{sec:snia}
\begin{figure}[ht!]
\centering
\includegraphics[width=8cm,trim={0.2cm 0.2cm 0.2cm 0.2cm},clip]{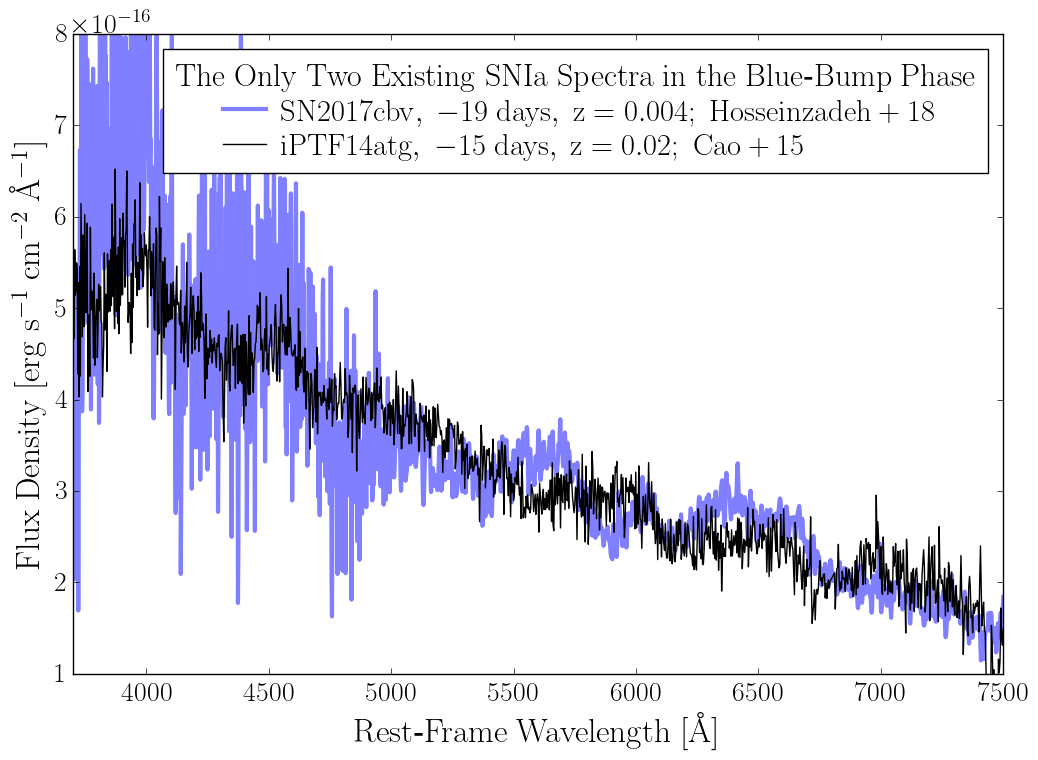}
\includegraphics[width=8cm,trim={0.2cm 0.2cm 0.2cm 0.3cm},clip]{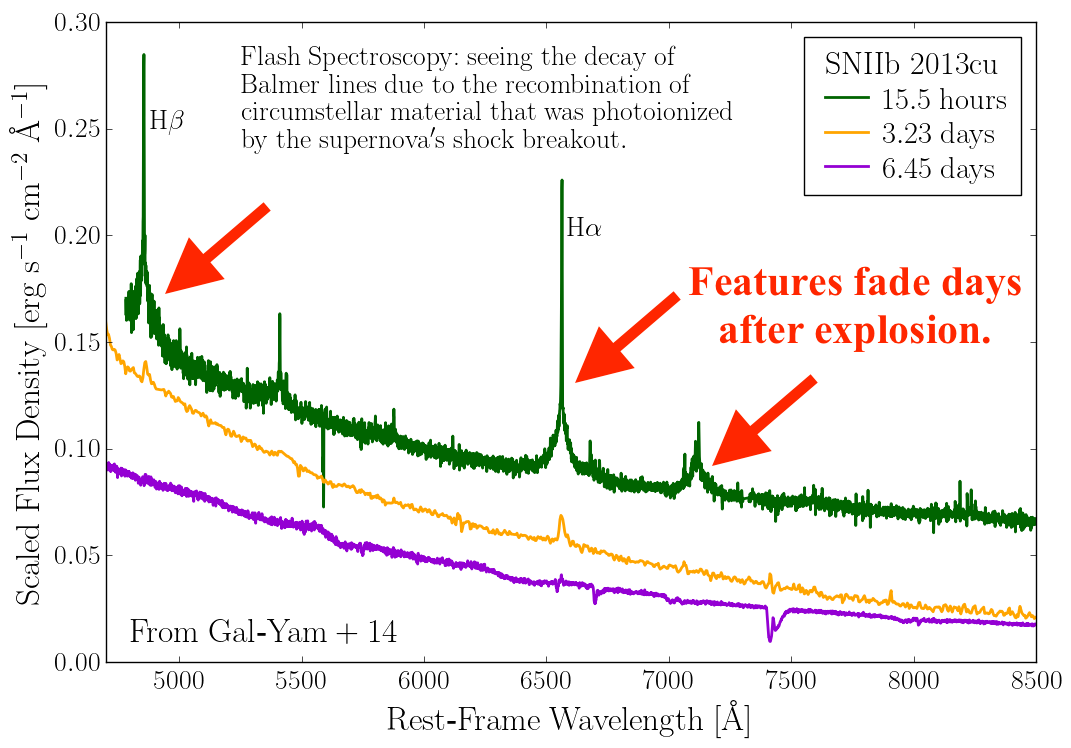}
\vspace{-5pt}
\caption{{\it Left:} The only two spectra of SNe\,Ia obtained during an early ``blue bump" (SN2017cbv from \citealt{2017ApJ...845L..11H}, blue, and iPTF14atg from \citealt{2015Natur.521..328C}, black). {\it Right:} Spectra exhibiting ``flash spectroscopy" for CC\,SN~2013cu (Type IIb; \citealt{2014Natur.509..471G}).}
\label{fig:2}
\end{figure}

\vspace{-5pt}
%%% MLG quick links
%%% SN2017cbv Hosseinzadeh+18 http://adsabs.harvard.edu/abs/2017ApJ...845L..11H
%%% iPTF14atg Cao+15 http://adsabs.harvard.edu/abs/2015Natur.521..328C
%%% SN2018oh  http://adsabs.harvard.edu/abs/2019ApJ...870...12L
Thermonuclear supernovae (Type Ia; SNe\,Ia) are used as standardizable candles but their explosion mechanism and the role of a binary companion remain unclear. The basic question is: does a white dwarf explode due to a merger with another white dwarf, or due to accretion from a non-degenerate companion? One of the clearest predicted signatures of the latter is a ``blue bump" in the light curve during the days after explosion, {\em if} the ejecta is shocked by a non-degenerate companion that lies along our line-of-sight \citep{2010ApJ...708.1025K}. This has so far been unambiguously observed for two SNe\,Ia, SN20018oh \citep{2019ApJ...870...13S,2019ApJ...870L...1D} and SN2017cbv \citep{2017ApJ...845L..11H}, and one peculiar SN\,Ia, iPTF14atg \citep{2015Natur.521..328C}. Spectra during the blue bump phase were obtained for the latter two (Fig. \ref{fig:2}) and appear similar to early spectra of ``normal" SNe\,Ia (e.g., SN2013dy, \citealt{2013ApJ...778L..15Z}). This similarity is curious, as is the fact that late-time spectroscopy did not reveal any of the expected circumstellar material (CSM) from a non-degenerate companion in the SN2017cbv system \citep{2018ApJ...863...24S}. {\it In the luminosity-timescale phase-space of Fig. \ref{fig:1}, such emission from shocked SN\,Ia ejecta occupies a region bounded by $\lesssim3$ days and $B\lesssim-17$ mag.} {\bf Building a sufficiently large sample of SNe\,Ia and establishing a robust connection between blue bumps at early times and non-degenerate companion stars will require LSST supported by ELT.}
% To build a sufficiently large sample to even {\em start} to understand the blue bump effect, and link it to the smoking gun of a non-degenerate's CSM, will be possible in the ELT+LSST era.
\vspace{-5pt}

%%%%%%%%%%%%%%%%%%%%%%%%%%%%%%%%%%%%%%%%%%%%%%%%%%%
\vspace{-15pt}
\section{Mass Loss Prior to Core Collapse}\label{sec:cc}
%%% MLG Quick Links
%%% Gal-Yam et al. (2014), SNIIb 2013cu http://adsabs.harvard.edu/abs/2014Natur.509..471G
\vspace{-5pt}
For SNe caused by the core collapse of massive stars ($>$8 M$_{\odot}$; CC\,SN), the high-energy emission from the shock breakout can ionize any surrounding CSM and the recombination produces transient, narrow lines
% superimposed on broad bases of $2500$ ${\rm km\ s^{-1}}$,
which fade in the week after explosion.
%; \citealt[e.g.,][]{2014Natur.509..471G,2016ApJ...818....3K}
Early-time spectra reveals the CSM's mass and distribution, which in turn constrains the late stages of mass loss from the progenitor star. In the case of SN\,IIb~2013cu (right panel of Fig. \ref{fig:2}), spectroscopy revealed a stellar wind which indicated that the progenitor was a Wolf-Rayet, an unstable luminous blue variable, or a yellow hypergiant undergoing an eruptive phase \citep{2014Natur.509..471G,2014A&A...572L..11G}.
% Additionally, very early-time spectroscopy during the post-shock breakout cooling phase provides unique constraints on the temperature, composition, and radius of the exploding star \citep{2009ApJ...702..226M,2016ApJ...832..139H}.
%For example, low-resolution spectra at $<2$ days after explosion revealed blackbody emission and the presence of doubly-ionized intermediate mass elements (O, C, N; \citealt{2009ApJ...702..226M}), and the rate of cooling can constrain the radius of the progenitor star \citep{2016ApJ...832..139H}.
{\it In the luminosity-timescale phase-space of Fig. \ref{fig:1}, early CC\,SN emission occupies a region bounded by $<3$ day and $R\lesssim-17$ mag.}
% very similar to the phase-space for ``blue bump" SNe\,Ia in \S~\ref{sec:snia}.
{\bf Increasing the number of CC\,SNe with flash spectroscopy and connecting the late stages of stellar evolution to explosions requires LSST+ELTs.} 

%%%%%%%%%%%%%%%%%%%%%%%%%%%%%%%%%%%%%%%%%%%%%%%%%%%
\vspace{-5pt}
\section{The Asymmetric Geometry of Stellar Explosions}\label{sec:spol}
\begin{figure}[ht!]
\centering
\includegraphics[height=6.5cm,trim={0.0cm 0.0cm 0.0cm 0.0cm},clip]{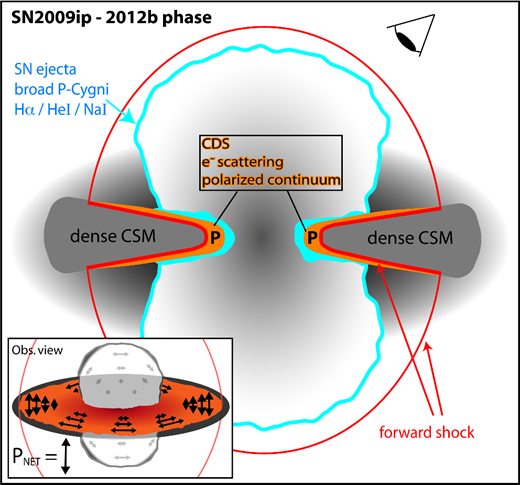}
\includegraphics[height=6.5cm,trim={0.0cm 0.5cm 0.9cm 0.9cm},clip]{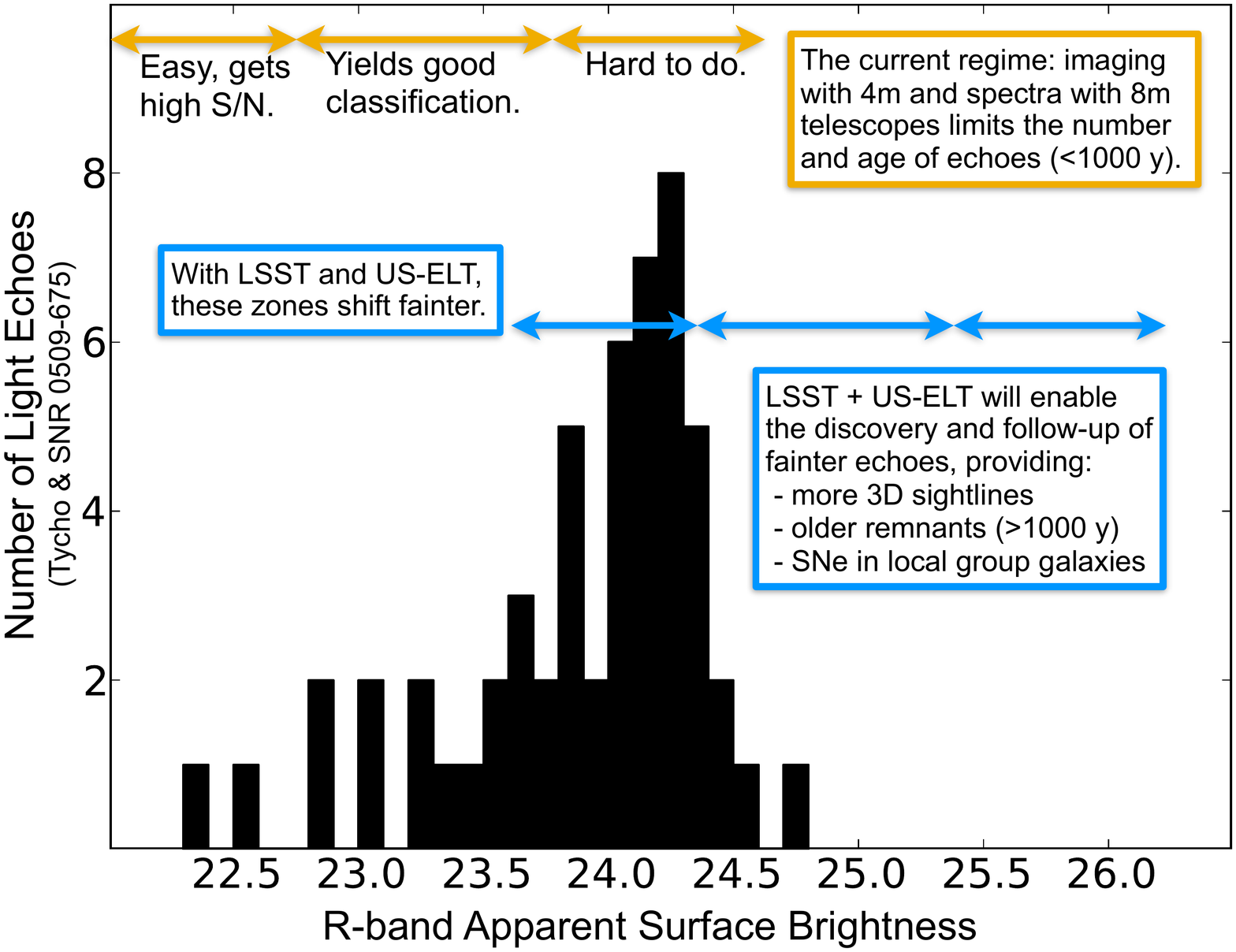}
\vspace{-5pt}
\caption{{\it Left:} This geometric asymmetry of bipolar ejecta lobes from a CC\,SN (cyan) impacting a toroid of CSM  ejecta (orange/red) was uniquely revealed for SN2009ip by spectropolarimetric data \citep{2014MNRAS.442.1166M}. {\it Right:} The surface brightness distribution of light echoes associated with SNR 0509-675 and Tycho \citep{Rest05,Rest08b}, showing current capabilities (orange) compared to that provided by ELTs and LSST (blue).  }
\label{fig:3}
\end{figure}

\vspace{-5pt}
SNe are complex three-dimensional structures, revealed by the shape of young SN remnants, asymmetries in nebular emission lines, and spectropolarimetry (SPol). Of the three, only SPol can yield a time-resolved history of the 3D structure of individual elements in the ejecta \citep{2008ARA&A..46..433W}, a direct signature of the explosion physics, progenitor star, and/or interaction with a companion star or CSM. SPol is useful for all SN types, and virtually all CC\,SNe reveal evidence of strong asymmetry \citep{2008ARA&A..46..433W}.
%, and Ca has been seen ejected at different angles and with different velocities than H, He, O, or Fe \citep{2009ApJ...705.1139M}.
For the controversial object SN2009ip ($25.5$ Mpc), SPol data demonstrated that the double-peaked light curve was caused by an asymmetric explosion impacting a toroidal distribution of nearby CSM (left panel of Fig. \ref{fig:3}; \citealt{2014MNRAS.442.1166M}).
% (see Fig. \ref{fig:3}). Once this asymmetry was accounted for, the total energy ($10^{51}$ $\rm erg$) provided the strongest proof that a true core collapse had occurred \citep{2014MNRAS.442.1166M}. SPol observations were possible with a 3.5m telescope only because SN2009ip was relatively nearby (25.5 Mpc).
{\it In the 2D phase-space diagram of Fig. \ref{fig:1}, SPol targets lie within populated regions but still represent a discovery frontier along a third dimension of measurement difficulty.} SPol is a rare capability among current facilities and, since the light is split not only by wavelength but also by polarization angle, a larger aperture telescope is required. Current data from 8-10m telescopes have given a framework to understand the basic SPol behavior of various classes of SNe, but the field remains photon starved and rich for exploration. {\bf The LSST will discover a plethora of explosive transients for which SPol could redefine our physical understanding, and we recommend that SPol instrumentation be developed for an ELT}.
%In the ELT era, the LSST will be the main discovery machine providing explosive transients that are suitable for SPol, especially the rare but valuable events such as SN2009ip which challenge our understanding. 
%An ELT system will change the study of spectropolarimetry from one of sparse data (a few events) to one with a substantial statistical sample (thousands of events).
\vspace{-5pt}

%%%%%%%%%%%%%%%%%%%%%%%%%%%%%%%%%%%%%%%%%%%%%%%%%%%
\vspace{-10pt}
\section{Light Echoes of Supernovae}\label{sec:echos}
\vspace{-5pt}
Most SNe are observed in their initial bright but optically thick phase, during which it is difficult to deduce the physical properties such as ejecta mass, kinetic energy, composition, and geometry. After thousands of years these properties can be studied in detail for the expanded SN remnants (SNR) in our Local Group. 
%, and sometimes also provide an identified compact object (as in the Crab pulsar).
However, it is nearly impossible to connect measurements from these two distinct phases because they are almost never observed for the same SN, with two exceptions: SN1987A in the LMC, and the few SNRs for which light echoes of the optically thick phase have been observed \citep[e.g.,][]{Rest08a,2012PASA...29..466R}. Furthermore, light echoes offer the unique ability to observe the same event from different lines of sight, because reflecting dust clouds offer different vantage points from which to see the explosion \citep[e.g.,][]{2011ApJ...732....3R,Sinnott13}. This provides a direct probe of the explosion's asymmetry which can otherwise only be traced indirectly with SPol for unresolved distant SNe (\S~\ref{sec:spol}).
% In favorable circumstances, even a spectroscopic time series of the event can be obtained \citep[e.g.,][]{Prieto14,Smith18}.
Light echoes also illuminate the surrounding material, offering insight to the progenitor's mass loss history (\S~\ref{sec:cc}). Since light echoes are slowly-evolving extended objects with very low surface brightness that often occur in crowded Galactic fields, our ability to discover them requires deep, wide-area, long-duration time-domain surveys with large-format CCDs (i.e., LSST). For example, the Mosaic imager on the 4m Mayall telescope was used to discover light echoes for Tycho's famous SN~1572 \citep{2008ApJ...681L..81R}, and follow-up spectroscopy with the 8.2m Subaru telescope revealed Tycho to be a Type Ia SN \citep{2008Natur.456..617K}. In the right panel of Fig. \ref{fig:3} we show a surface brightness histogram for light echoes around two SNRs, where the orange lines represent light echoes observable with 4m+8m facilities, and blue lines represent the future possibilities in the LSST+ELT era. {\it In the phase-space diagram of Fig. \ref{fig:1}, light echoes occupy the uncharted territory of timescales $>$200 days.} The LSST will provide more light echoes and more sightlines for a wider variety of SNe: there are $>$7 unsearched SNR in our Galaxy with an age of $<$2000 years (and future techniques may extend to 5000 years), many more light echoes in the LMC. % (where faint orphan light echoes have been discovered with HST).
%, and M31 remains uncharted territory.
{\bf LSST+ELTs will increase the number and diversity of SNe with echoes, enable the combination of physical measurements from SNRs with thousands of extragalactic SNe, and provide the critical missing link in understanding the end stages of stellar evolution.}

%%%%%%%%%%%%%%%%%%%%%%%%%%%%%%%%%%%%%%%%%%%%%%%%%%%
\vspace{-5pt}
\section{High-Redshift Transients}\label{sec:highz}
\vspace{-5pt}
The LSST+ELT revolution for high-$z$ transients could fill an entire white paper, so here we touch on just a few examples. {\it The phase-space diagram of Fig. \ref{fig:1} shows how the LSST's deep drilling fields will be able to detect SNe\,Ia out to $z<2$, and CC\,SNe to $z\lesssim1$.} {\bf The final LSST sample will contain millions of SNe, and only the ELTs can provide the required spectroscopy for the high-$z$ events and/or their host galaxies.} This extension of our SN\,Ia sample will allow us to study trends in their rest-frame ultraviolet spectra \citep[e.g.,][]{2012MNRAS.426.2359M}; disentangle the effects of progenitor metallicity and age on the explosion \citep[e.g.,][]{2009ApJ...691..661H}; constrain their distribution of delay-times (from birth to explosion; e.g., \citealt{2011MNRAS.417..916G}); and improve their use as cosmological standardizable candles. The sample of CC\,SNe to intermediate redshifts will enable a significantly more precise comparison between the volumetric rates of star formation and death \citep[e.g.,][]{2014ARA&A..52..415M}. The volume afforded by LSST+ELT's extension to high redshifts will also greatly increase the number of rare transients, such as intracluster SNe\,Ia, peculiar SNe, and super-luminous SNe.

%%%%%%%%%%%%%%%%%%%%%%%%%%%%%%%%%%%%%%%%%%%%%%%%%%%
\vspace{-5pt}
\section{Strongly Lensed Supernovae}\label{sec:lens}
\vspace{-5pt}
%%% MLG Quick Links
%%% Goldstein et al. (2018) http://adsabs.harvard.edu/abs/2017ApJ...834L...5G
%%% Kelly et al. (2018) http://adsabs.harvard.edu/abs/2015Sci...347.1123K
High-redshift SNe that are strongly gravitationally lensed by foreground galaxies and appear as multiple magnified point sources are extremely constraining for cosmological and dark matter studies. For events with significant time delays (days to weeks) the predicted reappearance could enable observations of very young SNe at high redshift, providing valuable information about their progenitor star (\S~\ref{sec:snia}). SN Refsdal ($z$=1.49) was the first to be clearly identified in {\it HST} imaging with four sources separated by $\theta{\sim}2$\arcsec\ \citep{2015Sci...347.1123K}. The second, SN2016geu ($z$=0.41), was discovered by iPTF and confirmed by {\it HST} to exhibit four sources with $\theta{\gtrsim}$0.3\arcsec\ \citep{2017Sci...356..291G}. \cite{2018arXiv180910147G} simulates that LSST will find ${\sim}380$ strongly lensed SNe per year, and that this sample will have a median redshift of $z{\sim}1$ and apparent brightness of $r{\sim}23.5$ mag, and typically exhibit two sources with $\theta{\sim}$0.6\arcsec. {\it In the phase-space diagram of Fig. \ref{fig:1}, most lensed SNe fall in with regular SNe, yet represent a discovery frontier due to their rarity and need for high-resolution imaging follow-up.} {\bf ELTs with adaptive optics will enable routine monitoring of gravitationally lensed SNe from LSST and significant extensions of their powerful cosmological utility} (e.g., the TMT Wide Field Optical Spectrograph will offer a resolution of 0.2\arcsec\ FWHM; \citealt{TMTArch}).

%%%%%%%%%%%%%%%%%%%%%%%%%%%%%%%%%%%%%%%%%%%%%%%%%%%
\vspace{-5pt}
\section{ELT \& LSST Synergy: Practical Aspects}\label{sec:datainfra}
\begin{figure}[ht!]
\centering
\includegraphics[width=8cm,trim={0 0 0 2.5cm},clip]{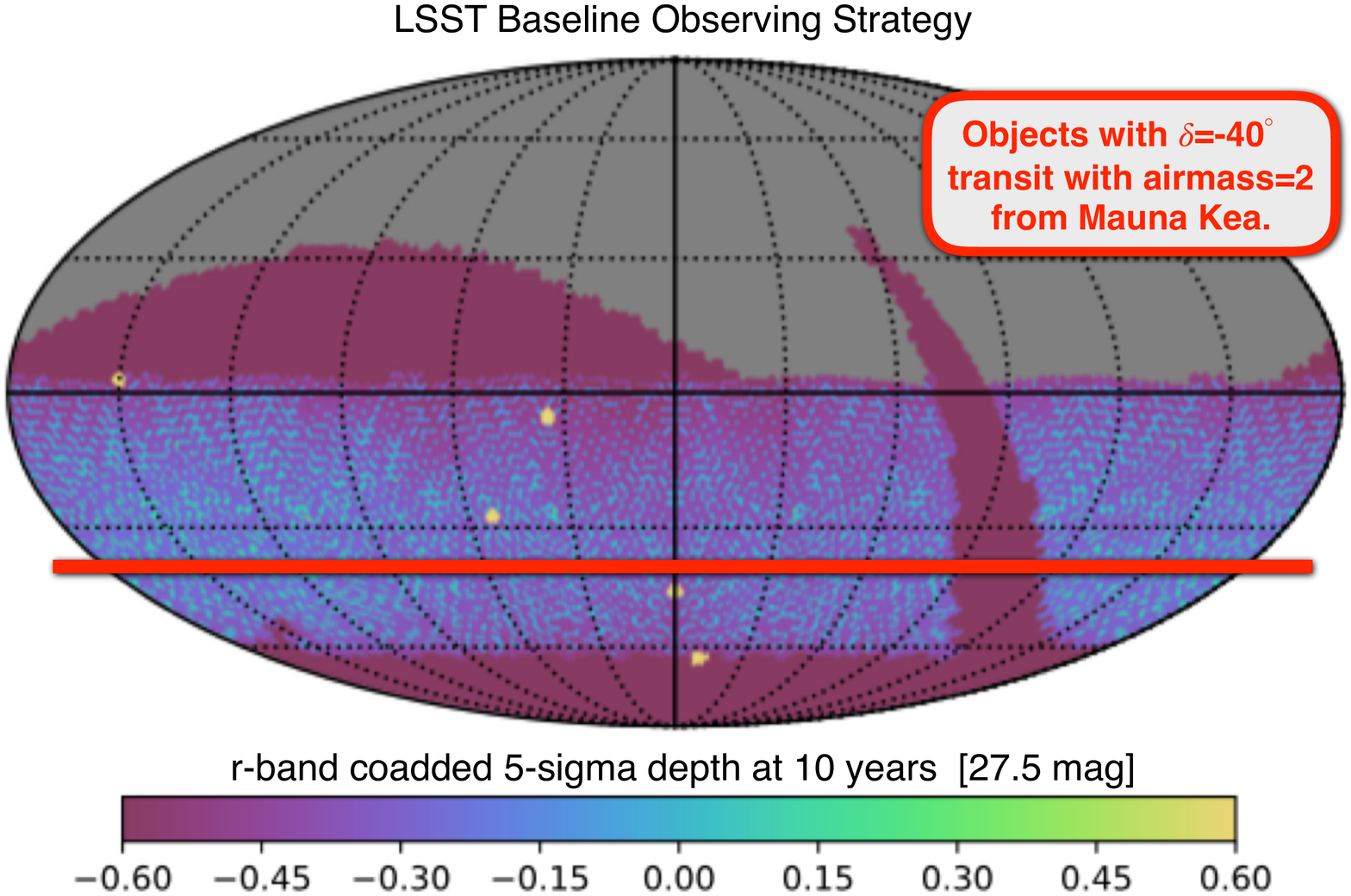}
\includegraphics[width=8cm,trim={0 0 0 2cm},clip]{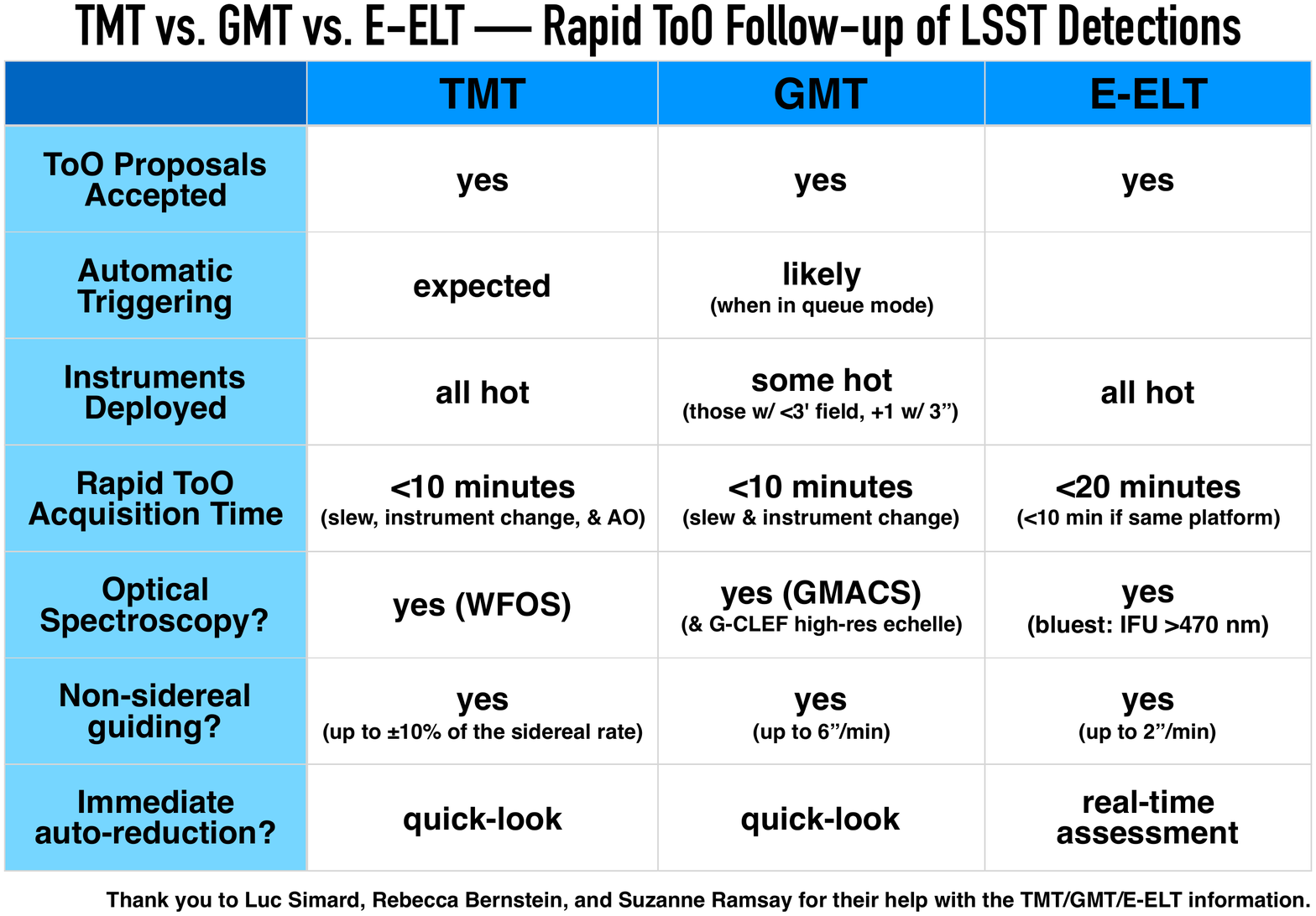}
\vspace{-5pt}
\caption{{\it Left:} Sky map of the LSST baseline survey to demonstrate significant northern (TMT) accessibility \citep{2008arXiv0805.2366I}. {\it Right:} A chart comparing the ToO follow-up of capabilities of the US and European ELTs \citep{GrahamTMT}.}
\label{fig:4}
\end{figure}

\vspace{-5pt}
{\bf Synergy with the ELTs will enable the LSST data to achieve its full scientific potential}, and this depends on their respective locations, designs, and timelines, as well as infrastructures for data management. While the European ELT (E-ELT) and the Giant Magellan Telescope (GMT) will be in the southern hemisphere with LSST, the TMT can access more than half the LSST baseline survey area from Mauna Kea (left panel of Fig. \ref{fig:4}). The LSST will begin operations in late 2023 and the ELTs aim for first light between 2025 and 2027, overlapping for at least half of the LSST main survey's 10 year duration \cite[e.g., slide 40 of][]{GrahamTMT}. In the era of ELTs and LSST, breakthrough time-domain science will require breakthrough technologies and infrastructures (e.g., \citealt{2019arXiv190305130O}). LSST will release data for transients within 1 minute of each new image \citep{2008arXiv0805.2366I,LSE-163}. These alerts will be ingested by community brokers, which will apply algorithms to find candidates of interest and are necessary to enable follow-up of, e.g., young SNe (\S~\ref{sec:snia},\ref{sec:cc}; \citealt{2018ApJS..236....9N,2018SPIE10707E..11S,BryanWP}). To obtain same-night spectra of LSST discoveries with an ELT requires ToO mode and constantly-mounted instruments; all three ELTs plan for these capabilities (right panel of Fig. \ref{fig:4}). As a final note, community efforts to maximize LSST discoveries in the discovery frontiers of Fig. \ref{fig:1} are underway \citep[e.g.,][]{2018arXiv181203146B}.
\vspace{-5pt}

%%%%%%%%%%%%%%%%%%%%%%%%%%%%%%%%%%%%%%%%%%%%%%%%%%%
% \section{Summary}

\pagebreak
%\textbf{References}
\let\oldbibliography\thebibliography
\renewcommand{\thebibliography}[1]{\oldbibliography{#1}
\setlength{\itemsep}{0pt}} %Reducing spacing in the bibliography.
\bibliography{main}

\end{document}